\documentclass[12pt]{article}
\usepackage[dvips]{graphicx}
\usepackage{epsfig,cite}
\usepackage {amsmath}
\usepackage{amssymb}
\usepackage{slashed}
\newcount\timecount
\newcount\hours \newcount\minutes  \newcount\temp \newcount\pmhours
\hours = \time
\divide\hours by 60
\temp = \hours
\multiply\temp by 60
\minutes = \time
\advance\minutes by -\temp
\def\hour{\the\hours}
\def\minute{\ifnum\minutes<10 0\the\minutes
            \else\the\minutes\fi}
\def\clock{
\ifnum\hours=0 12:\minute\ AM
\else\ifnum\hours<12 \hour:\minute\ AM
      \else\ifnum\hours=12 12:\minute\ PM
            \else\ifnum\hours>12
                 \pmhours=\hours
                 \advance\pmhours by -12
                 \the\pmhours:\minute\ PM
                 \fi
            \fi
      \fi
\fi
}

\def\monthname{\relax\ifcase\month 0/\or January\or February\or
   March\or April\or May\or June\or July\or August\or September\or
   October\or November\or December\else\number\month/\fi}

\def\bold#1{\setbox0=\hbox{$#1$}%
     \kern-.025em\copy0\kern-\wd0
     \kern.05em\copy0\kern-\wd0
     \kern-.025em\raise.0433em\box0 }

\def\beq{\begin{equation}}
\def\eeq{\end{equation}}

\def\ga{\mathrel{\raise.3ex\hbox{$>$\kern-.75em\lower1ex\hbox{$\sim$}}}}
\def\la{\mathrel{\raise.3ex\hbox{$<$\kern-.75em\lower1ex\hbox{$\sim$}}}}
\def\gev{{\rm \, Ge\kern-0.125em V}}
\def\tev{{\rm \, Te\kern-0.125em V}}
\def\gyr{{\rm \, G\kern-0.125em yr}}

\def\gappeq{\mathrel{\rlap {\raise.5ex\hbox{$>$}}
{\lower.5ex\hbox{$\sim$}}}}
\def\lappeq{\mathrel{\rlap{\raise.5ex\hbox{$<$}}
{\lower.5ex\hbox{$\sim$}}}}
\def\Toprel#1\over#2{\mathrel{\mathop{#2}\limits^{#1}}}

\def\m12{m_{1\!/2}}

\newcommand\iso[2]{\mbox{${}^{#2}${\rm #1}}}
\def\he#1{\iso{He}{#1}}
\def\be#1{\iso{Be}{#1}}
\def\li#1{\iso{Li}{#1}}

\def\bea{\begin{eqnarray}}
\def\eea{\end{eqnarray}}

\def\nrg{E}

\def\beqar{\begin{eqnarray}}
\def\eeqar{\end{eqnarray}}

\begin{document}

\begin{titlepage}
\pagestyle{empty}
\rightline{CERN-PH-TH/2010-163}
\rightline{UMN--TH--2911/10}
\rightline{FTPI--MINN--10/19}
\begin{center}
{\large {\bf Nuclear Reaction Uncertainties, Massive Gravitino Decays
and \\
\vskip 0.1in
the Cosmological Lithium Problem}}

\end{center}
\begin{center}
{\bf Richard~H.~Cyburt}$^{1}$, {\bf John~Ellis}$^2$,
{\bf Brian~D.~Fields}$^{3}$, \\
\vskip 0.1in
{\bf Feng Luo}$^{4}$, {\bf Keith~A.~Olive}$^{4,5}$,
and {\bf Vassilis~C.~Spanos}$^{6}$\\
\vskip 0.2in
{\small {\it
$^1${Joint Institute for Nuclear Astrophysics (JINA), National Superconducting
Cyclotron Laboratory (NSCL), Michigan State University, East Lansing,
MI 48824, USA}\\ 
$^2${TH Division, Physics Department, CERN, CH-1211 Geneva 23, Switzerland}\\
$^3${Center for Theoretical Astrophysics,
Departments of Astronomy and of Physics, \\ University of Illinois, Urbana, IL 61801, USA}\\
$^4${School of Physics and Astronomy, \\
University of Minnesota, Minneapolis, MN 55455, USA}\\
$^5${William I. Fine Theoretical Physics Institute, \\
University of Minnesota, Minneapolis, MN 55455, USA}\\
$^6${Institute of Nuclear Physics, NCSR ``Demokritos", GR-15310 Athens, Greece}} \\
}
\vskip 0.2in
{\bf Abstract}
\end{center}
{\small
We consider the effects of uncertainties in nuclear reaction rates on
the cosmological constraints on the decays of unstable particles
during or after Big-Bang nucleosynthesis (BBN). 
We identify the nuclear reactions due to non-thermal hadrons
that are the most important in perturbing
standard BBN, then quantify the uncertainties in these
reactions and in the resulting light-element abundances. 
These results also indicate the key nuclear processes for which
improved cross section data
 would allow different light-element abundances
to be determined more accurately, thereby making possible more precise probes of BBN
and evaluations of the cosmological constraints on unstable particles.
Applying this analysis to models with unstable gravitinos decaying
into neutralinos, we calculate the likelihood function for the light-element
abundances measured currently, taking into account the current experimental errors
in the determinations of the relevant nuclear reaction rates.
We find a region of the gravitino mass and abundance in which the
abundances of deuterium, \he4 and \li7 may be fit with $\chi^2 = 5.5$,
compared with $\chi^2 = 31.7$ if the effects of gravitino decays are unimportant.
The best-fit solution is improved to $\chi^2 \sim 2.0$ when the lithium abundance is taken 
from globular cluster data. Some such re-evaluation of the observed light-element 
abundances and/or nuclear reaction rates would be needed
if this region of gravitino parameters is to provide a complete solution to the cosmological
\li7 problem.}


\vfill
\end{titlepage}

\section{Introduction}

Late-decaying massive particles are generic features of plausible
extensions of the Standard Model, such as supersymmetry.
Cosmological constraints on such models are imposed by the observed
astrophysical abundances of the light elements~\cite{Lindley:1984bg}-\cite{jp}, 
which differ little from
those calculated in standard Big-Bang Nucleosynthesis (BBN)~\cite{cfo1,cfo2,bbn2,cyburt,cfo5}.
An accurate calculation of the constraints imposed by astrophysical
observations on the abundance of a late-decaying massive particle
requires taking into account not only the uncertainties in the
astrophysical observations but also the uncertainties in the
nuclear reaction rates that contribute to the production of light
elements in both standard and modified BBN scenarios.

We report in this paper on a study of the effects on the
astrophysical abundances of the light elements deuterium,
\he3, \he4, \li6 and \li7 of the uncertainties in the rates for 36
different nuclear reactions. Our central result can be expressed
as a $36 \times 5$ matrix that may be used, e.g., to calculate the
cumulative uncertainties in the BBN light-element abundances induced
by the uncertainties in any given set of nuclear data, and to estimate the changes in
the calculated abundances that would be induced by any updates
of the measurements of the nuclear reaction rates. In particular,
our analysis shows which uncertainties in reaction rates have the
greatest impact on the light-element abundances calculated in
standard BBN. Our analysis therefore bears upon the apparent
\li7 problem for standard BBN~\cite{cfo5}, and one of our main interests
is in the application of our analysis to the constraints on the modifications 
to standard BBN that occur in models with late-decaying particles.

Such particles appear, for example, in supersymmetric models with
gravity-mediated supersymmetry breaking. If the gravitino is the
lightest supersymmetric particle (LSP), then the next-to-lightest
supersymmetric particle (NLSP) is relatively long-lived~\cite{SFT,0404231,eoss5}. 
Alternatively,
if the lightest neutralino $\chi$ is the LSP, then the gravitino is long-lived.
In this paper we revisit the second possibility, as an illustration of the
incorporation of the effects
of the current uncertainties in the relevant nuclear reaction rates.

Neglecting these uncertainties, we analyzed previously~\cite{ceflos}
the constrained minimal supersymmetric extension of the Standard 
Model (CMSSM), in which the supersymmetry-breaking
masses for spartners of Standard Model particles are assumed to be
universal and the gravitino is assumed to be more massive~\cite{cmssmwmap}.
We found that there were strips in representative gravitino mass vs. abundance 
$(m_{3/2}, \zeta_{3/2})$ planes where the discrepancy between the
measured \li7 abundance and that calculated in standard BBN
could be reduced by the effects of late-decaying gravitinos without
destroying the successful BBN predictions for the other
light elements, particularly deuterium.

In this paper, as an application of our general analysis of the
implications of uncertainties in nuclear reaction rates, we
include them in a re-evaluation of this possible supersymmetric
solution to the \li7 problem. We re-calculate the global likelihood
function $\chi^2$ in the same representative $(m_{3/2}, \zeta_{3/2})$ 
planes, now including the uncertainties in the measured abundances
as well as the nuclear reaction rates. We confirm that $\chi^2$ is 
indeed minimized along the strips found previously, being reduced typically by
$\sim 26$ units. However, the quality of the best global fit to the
light-element abundances is still not very good: $\chi^2 = 5.5$ 
(for one effective degree of freedom - 
the three abundance measurements minus two fit parameters). 
Thus, if late-decaying
gravitinos are to solve the \li7 problem, they will need some help,
either from changes in some reaction rates outside the uncertainties
currently stated and/or changes in the measured deuterium and/or \li7 abundance,
and we discuss some such possibilities.

\section{Principal Nuclear Reaction Rates}
\label{sec:ratesandabund}

We first discuss the nuclear reactions included in our analysis.
The principle application of their non-thermal rates in the context of nucleosynthesis with 
late-decaying particles was discussed at length in \cite{ceflos}, so here we review only some 
important  details relevant to the current analysis. 

As described in \cite{ceflos}, 
non-thermal hadrons $h \in (p,n)$ are injected into the cosmological
baryon/photon plasma by the decay of a heavy particle such as the NLSP,
and then interact with the cosmic medium through which they travel.
Non-thermal particle propagation
is determined by competition among the various interactions
that lead to particle losses--continuous energy losses as
well as elastic and inelastic collisions.
These loss processes are always rapid compared to
the cosmic expansion rate. Thus, to a good approximation the
non-thermal particle spectra are set by an equilibrium
between injection and losses.  These propagated, equilibrium spectra
then determine the rates of non-thermal interactions with
light nuclei via convolution with the relevant cross sections.
That is, for the process $hb \rightarrow \ell$ of a non-thermal
hadron interacting with a thermal background species $b$ to
produce light element $\ell$, the interaction rate per target $b$ is
$\Gamma_{hb \rightarrow \ell} = \int N_h(\nrg) \ v \  \sigma_{hb \rightarrow \ell}(\nrg) \ d\nrg$,
where $N_h(\nrg)$ is the spectrum of non-thermal $h$ having
kinetic energy $\nrg$,
and $\sigma_{hb \rightarrow \ell}$ is the cross section
for the process at hand. The non-thermal processes  considered here are listed in Table~\ref{tab:reactions}.

The uncertainty in the reaction rate
$\Gamma_{hb \rightarrow \ell}$, due to 
cross-section errors $\delta \sigma_{hb \rightarrow \ell}$,  is
\beqar
\delta \Gamma_{hb \rightarrow \ell}
  & = & \int N_h(\nrg) \ v \  \delta \sigma_{hb \rightarrow \ell}(\nrg) 
          \ d\nrg \\
  & \equiv & \epsilon_{hb \rightarrow \ell} \ \Gamma_{hb \rightarrow \ell}
\eeqar
where $\epsilon_{hb \rightarrow \ell} 
  = \delta \Gamma_{hb \rightarrow \ell}/\Gamma_{hb \rightarrow \ell}$ 
characterizes
the fractional error in the rate.
The propagated non-thermal spectra  $N_h(\nrg)$
generally increase to a peak at $\nrg \sim few \ \rm GeV$, i.e.,
at energies far above thermal
energies, the Gamow peak, and any reaction threshold.  
Non-thermal rates, unlike thermal rates, are
sensitive to cross section behaviors over much larger
ranges of energies.  The cross sections
typically grow rapidly above threshold, and then in some cases
(fusion processes) drop strongly above $\nrg \sim few \times 10$ MeV,
or in other cases (spallation processes) remain nearly constant
or drop slowly at high energies. 
We should expect the uncertainties in
non-thermal rates often to be larger than the typical uncertainties
in the thermal rates, which are sensitive to a much narrower range
of energies around the  Gamow peak, 
typically $\sim 0.1-0.3$ MeV.

In principle, the rates $\Gamma_{hb \rightarrow \ell}$
have additional uncertainties due to those in the non-thermal
spectra $N_{h}$, which would in general need to be
added to the cross-section errors.
Since the $N_h$ are determined by sources and losses, they reflect
the uncertainties in these processes.   In practice, the dominant losses
are typically electromagnetic energy losses to the plasma, 
the rates for which are relatively well known.
Even in the regimes where the losses are dominated by scattering,
the relevant cross-section errors are better known than
typical reactions involving light elements.
The source (e.g., NLSP decay) spectra are also well-determined by supersymmetry and
Standard-Model physics.
Consequently, the errors in $N_h$ should be relatively small,
and thus the uncertainties in the reaction rates should be dominated
by the errors in the light-element cross sections that we have highlighted.

We have estimated uncertainties for the non-thermal reactions
by comparing nominal cross-section fits with 
experimental measurements.
The fitting functions $\sigma(\nrg)$ typically provide
good or excellent fits to the data.  However, the data
themselves are often sparse over the
large energy ranges of interest.  Unfortunately, this
paucity of data is particularly acute for the
spallation reactions $h \he4 \rightarrow (h,{}^{2}A,{}^{3}A)+\cdots$,
which are among the most important, as we shall see.
In each case, we estimate conservatively the typical 
fractional size $\epsilon = \delta \sigma/\sigma$ 
of the experimental error bars over the energies
where the cross section is substantial (i.e., near maxima
for strongly-peaked cross sections, and out to $\sim few$ GeV
for flat cross sections).  In the following section, we determine
which of these reactions have the most important impacts on
the light-element abundances, and we report  the uncertainties for those in Table \ref{tab:reactions}.

\begin{table}
\caption{Nuclear reactions of non-thermal particles, including the most important
of the estimated uncertainties in the cross sections.
\label{tab:reactions}}
\begin{center}
\begin{tabular}{cccl}
\hline\hline
Code & Reaction & Uncertainty $\epsilon$ & Reference \\
\hline
1 & $p\he4 \rightarrow d\he3$ &  & Meyer~\cite{meyer}\\
2 & $p\he4 \rightarrow np\he3$ & 20\% & Meyer~\cite{meyer}\\
3 & $p\he4 \rightarrow ddp$ & 40\% & Meyer~\cite{meyer}\\
4 & $p\he4 \rightarrow dnpp$ & 40\% & Meyer~\cite{meyer}\\
5 & $d\he4 \rightarrow \li6 \gamma$ &  & Mohr~\cite{mohr}\\ 
6 & $t\he4 \rightarrow \li6 n$ & 20\% & Cyburt et al.~\cite{cefo}\\
7 & $\he3\he4 \rightarrow \li6 p$ & 20\% & Cyburt et al.~\cite{cefo} \\
8 & $t\he4 \rightarrow \li7 \gamma$ &  & Cyburt~\cite{cyburt} \\
9 & $\he3\he4 \rightarrow \be7 \gamma$ &  & Cyburt and Davids~\cite{cd} \\
10 & $p\li6 \rightarrow \he3\he4$ &  & Cyburt et al.~\cite{cefo} \\
11 & $n\li6 \rightarrow t\he4$ &  &  Cyburt et al.~\cite{cefo} \\
12 & $pn \rightarrow d \gamma$ &  & Ando, Cyburt, Hong, and Hyun~\cite{ando} \\
13 & $pd \rightarrow \he3 \gamma$ &  & Cyburt et al.~\cite{cefo} \\
14 & $pt \rightarrow n \he3$ &  & Cyburt~\cite{cyburt} \\ 
15 & $p\li6 \rightarrow \be7 \gamma$ &  & Cyburt et al.~\cite{cefo} \\ 
16 & $p\li7 \rightarrow \be8 \gamma$ &  & Cyburt et al.~\cite{cefo} \\
17 & $p \be7 \rightarrow \iso{B}{8} \gamma$ &  & Cyburt et al.~\cite{ceflos} \\
18 & $n p \rightarrow d \gamma$ &  & Ando, Cyburt, Hong, and Hyun~\cite{ando} \\
19 & $nd \rightarrow t \gamma$ &  & Cyburt et al.~\cite{cefo}\\ 
20 & $n\he4 \rightarrow dt$ &  & Meyer~\cite{meyer} \\
21 & $n\he4 \rightarrow npt$ & 20\% & Meyer~\cite{meyer}\\
22 & $n\he4 \rightarrow ddn$ & 40\% & Meyer~\cite{meyer}\\
23 & $n\he4 \rightarrow dnnp$ & 40\% & Meyer~\cite{meyer}\\
24 & $n \li6 \rightarrow \li7 \gamma$ &  & Cyburt et al.~\cite{cefo} \\
25 & $n$ (thermal) &  & --- \\
26 & $n \be7 \rightarrow p \li7$ &  & Cyburt et al.~\cite{cefo} \\
27 & $n \be7 \rightarrow \he4 \he4$ &  & Cyburt et al.~\cite{ceflos} \\
28 & $p \li7 \rightarrow \he4 \he4$ &  & Cyburt et al.~\cite{cefo} \\
29 & $n \pi^+ \rightarrow p \pi^0$ &  & Meyer~\cite{meyer} \\
30 & $p \pi^- \rightarrow n \pi^0$ &  & Meyer~\cite{meyer} \\
31 & $p \he4 \rightarrow p p t$ & 20\% & Meyer ~\cite{meyer}\\
32 & $n \he4 \rightarrow n n \he3$ & 20\% & Meyer ~\cite{meyer} \\
33 & $n\he4 \rightarrow nnnpp$ &  & Meyer~\cite{meyer}\\
34 & $p\he4 \rightarrow nnppp$ &  & Meyer~\cite{meyer}\\
35 &$p\he4 \rightarrow N\he4 \pi$ &  & Meyer~\cite{meyer}\\
36 &$n\he4 \rightarrow N\he4 \pi$ &  & Meyer~\cite{meyer} \\
\hline\hline
\end{tabular}
\end{center}
\end{table}

For a sufficiently large abundance of gravitinos, the standard BBN predictions
are modified, and the resulting light-element abundances need to be compared with
observational determinations.
In~\cite{ceflos}, we used the abundances (or abundance ratios)
of D, \he4, \li7, \he3/D, and \li6/\li7 to determine the allowed
regions of parameter space defined by the gravitino mass, $m_{3/2}$, 
the gaugino mass, $m_{1/2}$, the ratio of Higgs vacuum expectation
values, $\tan \beta$, and the gravitino abundance, $\zeta_{3/2}$, characterized by
\beq
\label{eq:zeta}
\zeta_{3/2} \equiv \frac{m_{3/2} n_{3/2}}{n_\gamma}
  = m_{3/2} Y_{3/2} \eta ,
\eeq
where $n_{3/2}$ is the gravitino number density, $ Y_{3/2} = n_{3/2}/n_B$, and 
$ \eta = 6.19 \times 10^{-10}$ is the baryon-to-photon ratio 
from WMAP year 7 \cite{wmap7}. For our present
$\chi^2$ analysis, we restrict our attention to the elements that have 
definite observational abundances with which we can make a comparison,
namely the following.

\underline{D/H}:
We use the deuterium abundance as determined in several high-redshift
quasar absorption systems, which have a weighted mean abundance~\cite{bt98a,bt98b,omeara,pettini,kirkman,omeara2,pettini2}
\beq
\label{eq:D}
\left( \frac{\rm D}{\rm H} \right)_p = \left( 2.82 \pm 0.21 \right)
  \times10^{-5} ;
\eeq
where the uncertainty includes a scale factor of  1.7 due to the dispersion
found in these observations.
Since the D/H ratio shows considerable scatter,
it is likely that systematic errors dominate the uncertainties.
In this case it may be more
appropriate to derive the uncertainty using sample variance (see
e.g. \cite{cfo1}) which gives 
a more conservative range  D/H =
$(2.82 \pm 0.53) \times 10^{-5}$.  We comment further on this below.
The standard BBN result for D/H at the WMAP value for $\eta$ is $(2.52 \pm 0.17) \times 10^{-5}$,
showing potentially a slight discrepancy with the observed value, unless one adopts the larger uncertainty.

\underline{\he4}:
The \he4 abundance is determined from observations of 
extragalactic H II regions.  These abundance determinations are known to 
suffer from large systematic uncertainties~\cite{osk}. A recent analysis found~\cite{aos}
\beq
Y_p = 0.256\pm0.011 ,
\label{yp}
\eeq
and a similar central value was found in \cite{it}.
The standard BBN result for $Y_p$ at the WMAP value for $\eta$ is $0.2487 \pm 0.0002$, which
is consistent with observations, given the error in (\ref{yp}).

\underline{\li7/H}:
The \li7 abundance is derived from observations of low-metallicity halo 
dwarf stars.  Some $\ga 100$ such
stars show a plateau\cite{spite}
in (elemental) lithium versus metallicity,
with a small scatter consistent with observational uncertainties.
An analysis \cite{rbofn}
of field halo stars gives a plateau abundance of
\beq
\label{eq:li_halo}
\left( \frac{\rm Li}{{\rm H}} \right)_{\rm halo \star}
 = (1.23^{+0.34}_{-0.16}) \times 10^{-10} ,
\eeq
where the errors include
both statistical and systematic uncertainties.
As in the case of \he4, the errors are dominated by systematic uncertainties.
For example,
the lithium abundance in several globular clusters, tends to be somewhat higher~\cite{liglob,pasquini,thevenin,li_m92,lind,new}, and
we make some comparisons below to the result found in \cite{new}
of \li7/H = $(2.34 \pm 0.05) \times 10^{-10}$. However,
the standard BBN result for \li7/H at the WMAP value for $\eta$ is $(5.12 {}^{+0.71}_{-0.62})
 \times 10^{-10}$,
which differs significantly from the observed value, hence the \li7 problem \cite{cfo5}.
Note that the central values for the BBN abundances used here differ slightly from those
in \cite{cfo5}, primarily due to the small shift in $\eta$ as reported in \cite{wmap7}.

Recently, there have been several analyses which indicate that the \li7 abundance at
low metallicity falls below the typical plateau value and/or shows a significant amount
of dispersion \cite{asp06,hos,aoki,sbordone,asp10}. While these observations
apparently provide the first indications of Li depletion in metal-poor stars, it would appear
that it is operative only at extremely low metallicity,
[Fe/H] $\la -3$, whatever the particular depletion mechanism may be,
whether in the star or in the medium
prior to the star's formation. There is no observational evidence of any depletion 
at higher metallicity ($-3 \la$ [Fe/H] $\la -1.5$) from the standard BBN result
to the plateau value \cite{aoki,sbordone}, in contrast to the claim 
of~\cite{asp10}~\footnote{It was argued in \cite{asp10} that there are two plateau values
corresponding to [Fe/H] above and below -2.5.  However, the evidence for this assertion
is not convincing, as these data can be fit
equally well with a linear increase in $\log$Li vs. [Fe/H] as in~\cite{rbn,rbofn,hos,sbordone}.
This would point to a {\em lower} primordial Li abundance and a more severe problem
with respect to standard BBN predictions.}. 

To obtain our $\chi^2$ distribution, we combine the standard BBN uncertainties with
the observational errors in quadrature.  In the case of \li7, where the reported errors are
uneven, we use the upper error bar on the observation, and the lower error bar on the theory,
as we are interested in the region between these two central values.
Correspondingly, the likelihood function that we calculate is
\beq
\chi^2 \; \equiv \; \left(\frac{Y_p - 0.256}{0.011} \right)^2 + \left(\frac{\frac{D}{H} - 2.82 \times 10^{-5}}{0.27 \times 10^{-5}}\right)^2 +
 \left(\frac{\frac{\li7}{H} - 1.23 \times 10^{-10}}{0.71 \times 10^{-10}}\right)^2
 + \sum_i s_i^2 ,
 \label{chi2}
 \eeq
 where the $s_i$ are the  contributions to the total $\chi^2$ due to the nuisance parameters
 associated with varying one or more of the rates listed in Table \ref{tab:reactions}.
 Standard BBN has a large total $\chi^2 = 31.7$, primarily due to the discrepancy 
 in \li7.
There is a contribution  of  $\Delta \chi^2 \sim 30$ from the \li7
abundance, $\Delta \chi^2 \sim 1.2$ from the D/H abundance, and a smaller
contribution from \he4, corresponding to a $\sim 5-\sigma$ discrepancy 
overall~\footnote{We find that $\chi^2 = 21.8$ even when the globular cluster 
value of \li7/H is used,
 corresponding to a $4-\sigma$ effect.}.

Our treatments of the hadronic and electromagnetic components of the
showers induced by heavy-particle decays follow those in~\cite{ceflos}.
Also, we follow the calculations of decay branching ratios and particle
spectra described in~\cite{ceflos}. The only differences here are in the
nuclear reaction rates and their uncertainties that were discussed above.

\begin{figure}[ht!]
\begin{center}
\epsfig{file=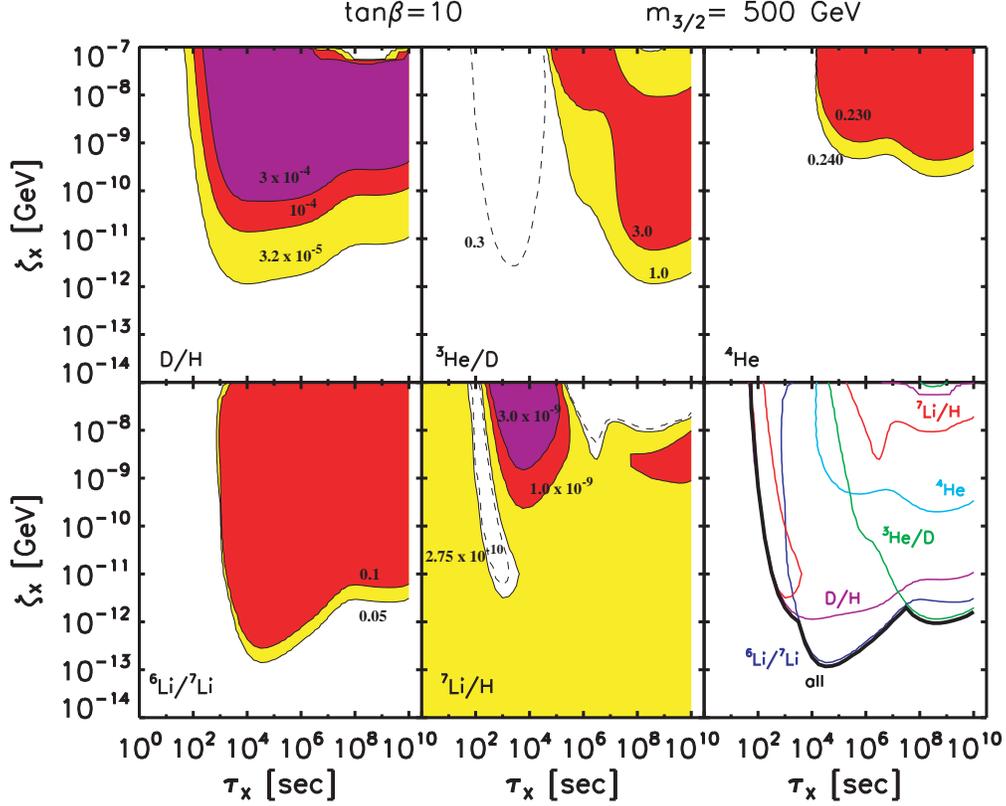,width=0.8\textwidth}
\end{center}
\caption{
\it 
Plots of the effects on the light-element abundances 
of the decays of a generic metastable particle
$X$ with a lifetime $\tau_X \in (1, 10^{10})$~sec,
using the decay spectra calculated for
$(m_{1/2},m_{3/2},\tan \beta) = (300~{\rm GeV}, 500~{\rm GeV}, 10)$,
in which case $B_p \approx 0.2$
and the electromagnetic branching rate is $B_{\rm EM} m_{3/2} = 115$ {\rm GeV}.
The $X$ abundance before decay is given by $\zeta_X = m_X n_X/n_\gamma$.
The white regions in each panel are those allowed
at face value by the ranges of the light-element abundances
reviewed in Section~\ref{sec:ratesandabund},
whilst the yellow, red and magenta regions correspond to
progressively larger deviations from the central values of the abundances.
\label{fig:new-zeta-tau}}
\end{figure}

We display in Fig.~\ref{fig:new-zeta-tau} the effects on the abundances of the
light elements deuterium, \he3, \he4, \li6 and \li7 of the decays of a generic
metastable particle $X$ with lifetime $\tau_X \in (1, 10^{10})$~sec. For
illustration, we assume the decay spectra calculated in~\cite{ceflos}
for the choice $(m_{1/2},m_{3/2},\tan \beta) = (300~{\rm GeV}, 500~{\rm GeV}, 10)$,
in which case the proton branching ratio $B_p \approx 0.2$
and the electromagnetic branching rate is $B_{\rm EM} m_{3/2} = 115$ {\rm GeV}.
In this figure we assume the nominal central values of the nuclear reaction rates
discussed in the text, and this figure may be compared directly
with Fig.~6 of~\cite{ceflos}. The main differences are in the upper left panel,
where the region where the deuterium abundance lies within the favoured
range is now pushed to values of $\zeta_X$ that are lower by a factor of about 2 
when $\tau_X < 10^6$~sec as compared with the results of \cite{ceflos}, and in the
lower middle panel, where the region of acceptable \li7 abundance extends
to lower $\zeta_X$ when $\tau_X \sim 10^3$~sec. Both these effects are due to
the inclusion of the reactions $n \he4 \to n n \he3$ and $p \he4 \to p p t$,
and have the effect of pushing the location of a possible `solution' of the \li7
problem also to lower $\zeta_X$.

\begin{figure}[ht!]
\begin{center}
\epsfig{file=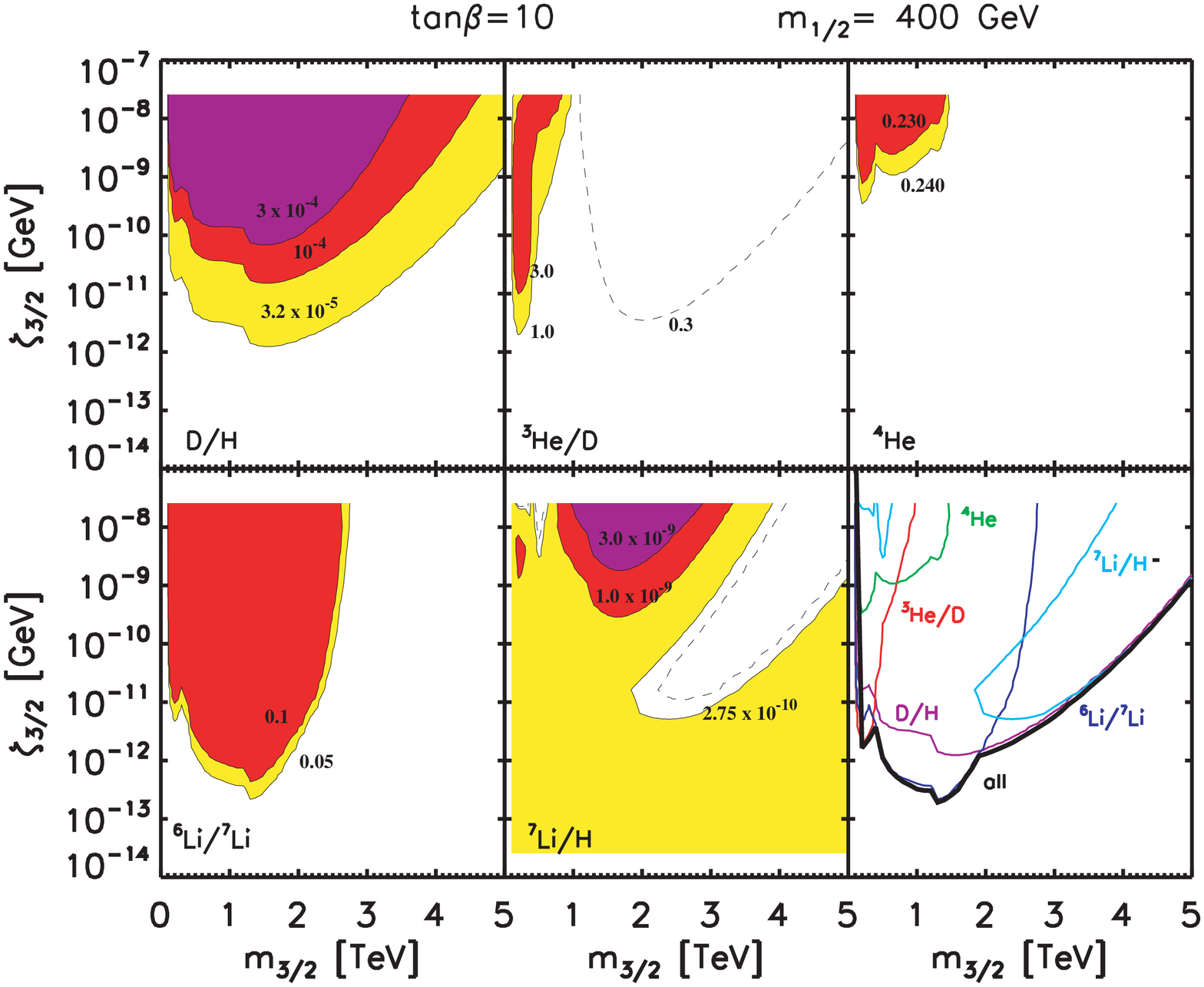,width=0.8\textwidth}
\end{center}
\caption{\it
The effects of the decays of a gravitino with variable mass $m_{3/2}$
on the different light-element abundances for a specific point 
(benchmark C of \protect~\cite{bench}) with
$m_{1/2} = 400$~GeV on the WMAP coannihilation strip for a CMSSM scenario
with $\tan \beta = 10, A_0 = 0$. As in the previous figure,
the white regions in each panel are those allowed
at face value by the light-element abundances
reviewed in Section~\ref{sec:ratesandabund}, and the yellow, red, and magenta regions 
correspond to
progressively larger deviations from the central values of the abundances.
}
\label{fig:elementsCm32}
\end{figure}

We show in Fig. ~\ref{fig:elementsCm32}, one generic $(m_{3/2}, \zeta_{3/2})$ plane, 
also without the inclusion of uncertainties in the non-thermal rates in Table \ref{tab:reactions}. 
This plot is based on a specific CMSSM point (benchmark C of~\cite{bench})
with $m_{1/2} = 400$ GeV, $A_0 = 0$, and $\tan \beta = 10$.  The universal scalar mass
is set to $m_0 = 90$ GeV to get the correct WMAP density for dark matter.
The lightest neutralino mass is about 165 GeV for this point, and for gravitino masses
larger than this we have neutralino dark matter with an unstable massive gravitino.
As in Fig. \ref{fig:new-zeta-tau},  the region where the deuterium abundance lies within the favoured range is now also pushed to lower $\zeta_{3/2}$ when $m_{3/2} \la 2$ TeV as compared
to the results in \cite{ceflos}, and the region of acceptable \li7 abundance extends
to lower $\zeta_{3/2}$ when $m_{3/2}$ is between 2 -- 3 TeV. 
In the lower right panel, we see marginal compatibility between the \li7 constraint (light blue)
and the other constraints for $m_{3/2} \ga 3$~TeV. This region will be the focus of our discussion
in the following $\chi^2$ analysis.

\section{Incorporation of Uncertainties}

Of the 36 interactions we study, there are just 10 whose uncertainties induce
non-negligible uncertainties in the light-element abundances, namely the reactions
2, 3, 4, 6, 7, 21, 22, 23, 31, and 32 in Table~\ref{tab:reactions}. 
Their uncertainties are not important for the \he4
abundance $Y_p$, but are potentially important for the deuterium, \he3, \li6 and \li7
abundances. 

To explore the effect of reaction uncertainties,
we quantify the reaction sensitivity as follows.
Using the set of unperturbed non-thermal reaction rates $\{ \Gamma_i^0 \}$,
we find the unperturbed abundances of light elements: examples of
these results appear in
Figures \ref{fig:new-zeta-tau} and \ref{fig:elementsCm32}.  
For a given light element $\ell$, we call the
unperturbed abundance $y_\ell^0$.
Then, for any single reaction $j$, we consider changes in the rate by a factor
$1+\epsilon$: $\Gamma_j^\prime = (1+\epsilon) \Gamma_j^0$, leaving all other non-thermal
(and thermal) rates unchanged. We evaluate the new resulting light-element abundances 
for a wide range of values for $\epsilon$ including both positive and 
negative values and write the new, perturbed $\ell$ abundance
as $y_\ell^\prime|_{{\rm rxn}j} \equiv y_\ell^0 + \delta y_\ell|_{{\rm rxn}j}$.
In this way, we were able to identify the 10 reactions listed above as potentially playing
an important role in altering the light-element abundances.
Our final results are based on a Gaussian distribution
of rates with widths give by the values of $\epsilon$
chosen according to the uncertainty
estimates in Table \ref{tab:reactions}.

\begin{figure}
\begin{center}
\epsfig{file=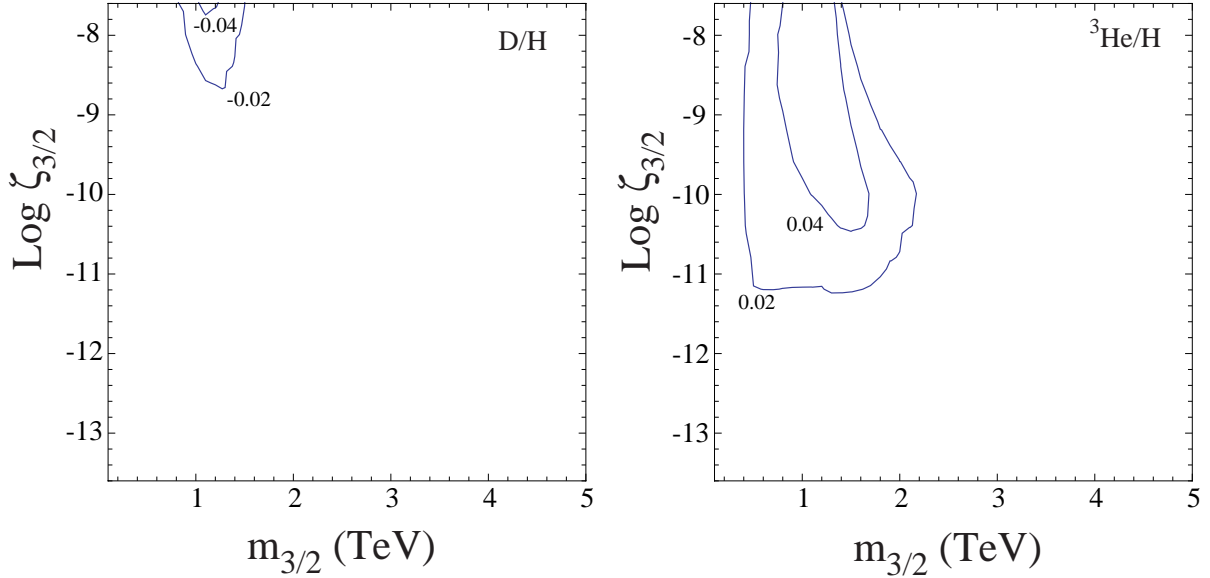,height=8cm}
\end{center}
\caption{\it
The effects in the $(m_{3/2}, \zeta_{3/2})$ plane of the 20\% uncertainty in the
rate for the reaction 2 ($p\he4 \rightarrow np\he3$) on the abundances of
deuterium (left) and \he3 (right). Contours show the relative changes in the light-element
abundances.}
\label{fig:reaction2}
\end{figure}

\begin{figure}
\begin{center}
\epsfig{file=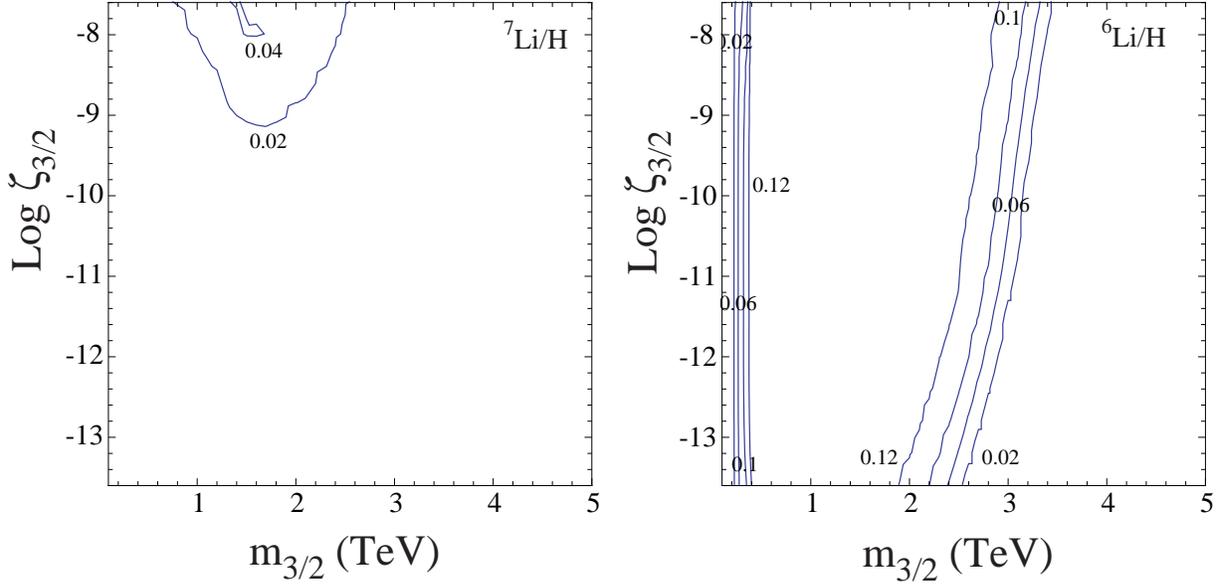,height=8cm}
\end{center}
\caption{\it
Similar to Fig.~\protect\ref{fig:reaction2}, for the reaction 6 ($t\he4 \rightarrow \li6n$),
in this case showing  the effect on \li7/H (left) and \li6/H (right).}
\label{fig:reaction6}
\end{figure}

\begin{figure}
\begin{center}
\epsfig{file=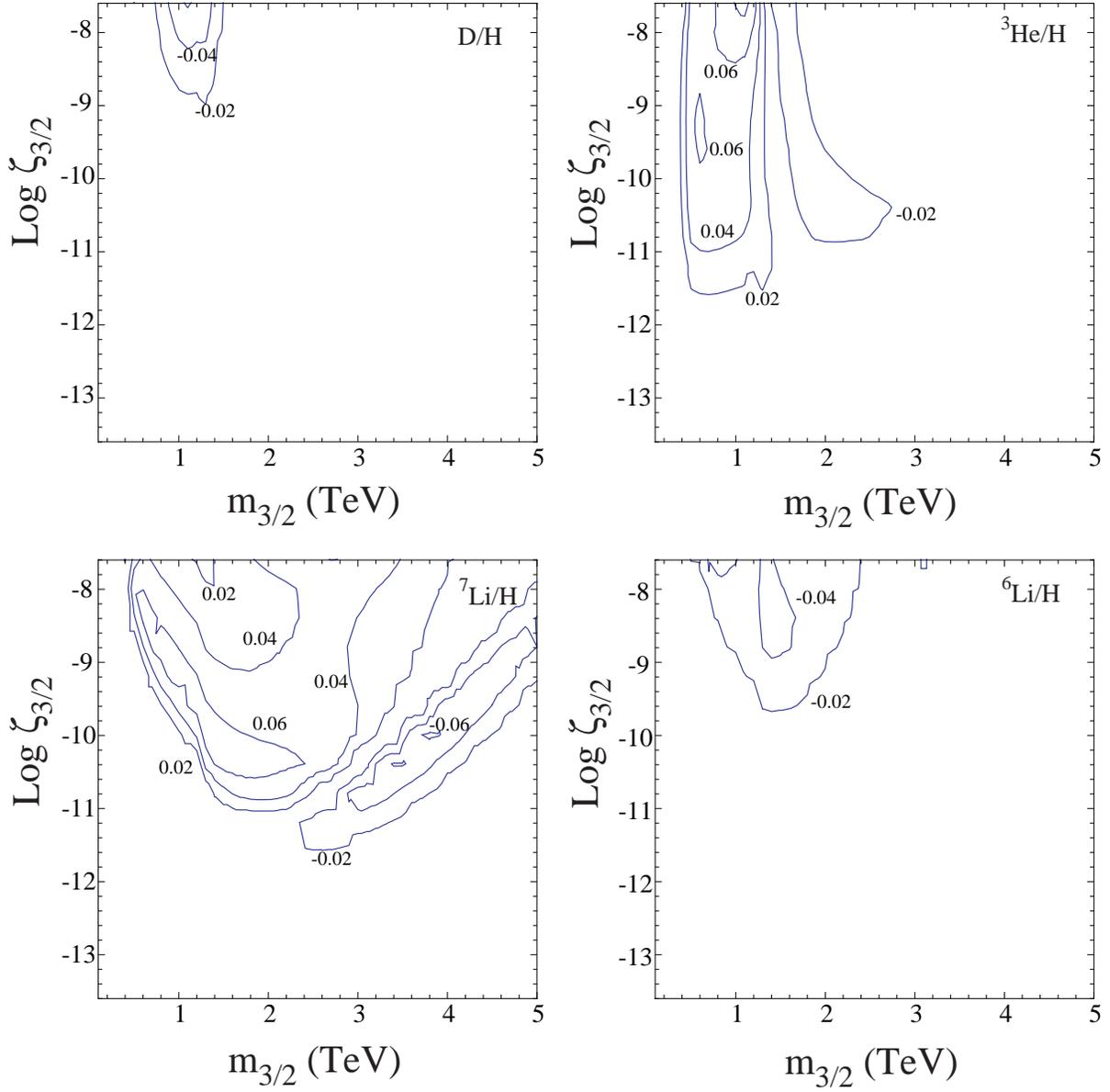,height=16cm}
\end{center}
\caption{\it
Similar to Fig.~\protect\ref{fig:reaction2}, for the reaction 21 ($n\he4 \rightarrow npt$),
showing the effects on all four light elements deuterium (upper left), \he3 (upper right), 
\li7 (lower left) and \li6 (lower right).}
\label{fig:reaction21}
\end{figure}

We display in Figs.~\ref{fig:reaction2},  
\ref{fig:reaction6}, \ref{fig:reaction21},  and \ref{fig:reaction23} the
effects of the uncertainties in these reaction rates on the abundances of
each of the key elements among deuterium, \he3, \li6 and \li7, 
each in a $(m_{3/2}, \zeta_{3/2})$ plane.
We concentrate here on benchmark point C, as the effect of perturbing the interactions
is qualitatively similar for the other benchmark points we consider below.

For example,
in Fig.~\ref{fig:reaction2} we show the effect of the $p\he4 \rightarrow np\he3$
(reaction 2) in Table \ref{tab:reactions} on the abundances of D/H (left) and \he3/H (right).
In this case the effect on \li6 and \li7 is negligible.  We plot contours showing
\beq
\left. \frac{\delta y_\ell}{y_\ell}\right|_{{\rm rxn}j} \equiv 
  \frac{y_{\ell}^\prime|_{{\rm rxn}j} - y_\ell^0}{ y_\ell^0} \ ,
\eeq
the
relative change in the light-element abundance when rate $j$ is perturbed by a factor
$(1+\epsilon)$.
For reaction 2, we estimate a 20\% uncertainty in the rate and,  
as one can see, the effect on D/H is always less than 4\% and occurs at
very high gravitino abundances.  The effect on \he3 is also relatively small, but extends over a larger portion of the parameter space.
The effect of reaction 31( $p \he4 \rightarrow p p t$) is qualitatively similar to the one shown
in Fig.~\ref{fig:reaction2}.

For reactions 3 and 4, corresponding to  $p\he4 \rightarrow ddp$ and  $p\he4 \rightarrow dnpp$
respectively, 
we estimate an uncertainty of 40\%. However, even with the larger uncertainty,
the relative change in D/H for both rates is still less than 4\%, 
extends down to lower $\zeta_{3/2} \sim 10^{-10}$~GeV, 
and the \he3 abundance variation is even smaller.
Accordingly, we do not show these examples.

In Fig.~\ref{fig:reaction6}, the effects of reaction 6 corresponding to
$t\he4 \rightarrow \li6 n$ are displayed (the effects of reaction 7 corresponding
to $\he3\he4 \rightarrow \li6 p$ are similar but weaker by a factor of 2):
we estimate 20\% uncertainties for these reactions.
We see that in this case, while there is some effect on the abundance of \li7,
the dominant effect of varying this rate is on \li6, where changes can be as large as
12\% for almost all the values of $\zeta_{3/2}$ shown when $m_{3/2} \sim 1$~TeV.  
We find similar results for points E and L, whilst for point M (see below) similarly large changes
in \li6 are centered around $m_{3/2} \sim 2$ TeV. 
The effects on deuterium and \he3 are negligible for reactions 6 and 7.

The effects of reaction 21 ($n\he4 \rightarrow npt$), for which we also estimate an
uncertainty of 20\%, on all four light elements are shown in Fig.~\ref{fig:reaction21}. 
The possible effect on \li7 is largest, amounting possibly to
a reduction in the \li7 abundance by up to 
6\% in a diagonal region extending from $m_{3/2} \ga 3 $ TeV. 
(For benchmark point M,
the reduction in the \li7 abundance occurs at $m_{3/2} \ga 4 $ TeV.) 
Reaction 32 ($n \he4 \rightarrow n n \he3$) 
shows effects somewhat similar (though in general a bit weaker) to those
seen in Fig.~\ref{fig:reaction21}, and is not shown separately.

\begin{figure}
\begin{center}
\epsfig{file=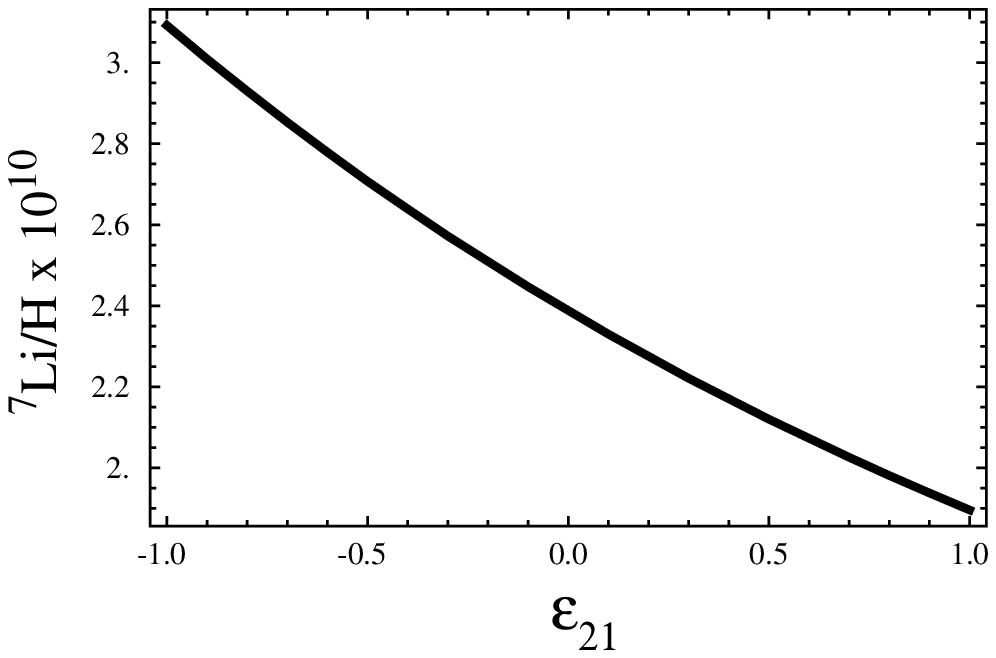,height=8cm}
\epsfig{file=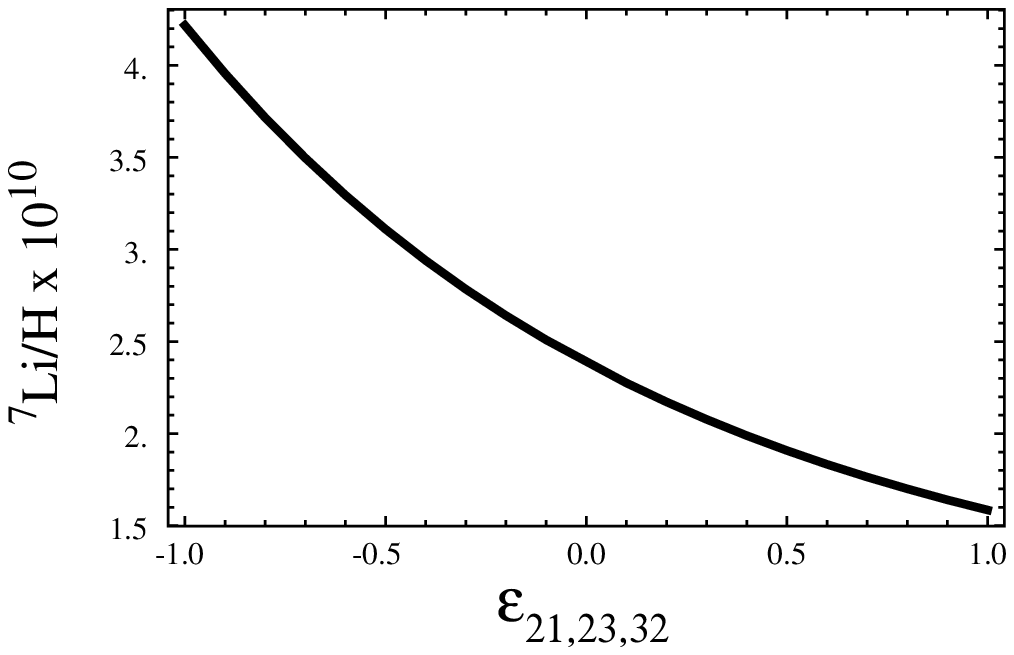,height=8cm}
\end{center}
\caption{\it
The \li7 abundance as a function of $\epsilon$: $\epsilon = -1$ is equivalent 
to turning off the rate, and $\epsilon = 0$ leaves the rate unperturbed.  In the upper panel
we show the effect of reaction 21 alone, and in the lower panel we show the combined effect of 
reactions 21, 23, and 32.}
\label{fig:livseps}
\end{figure}

\begin{figure}
\begin{center}
\epsfig{file=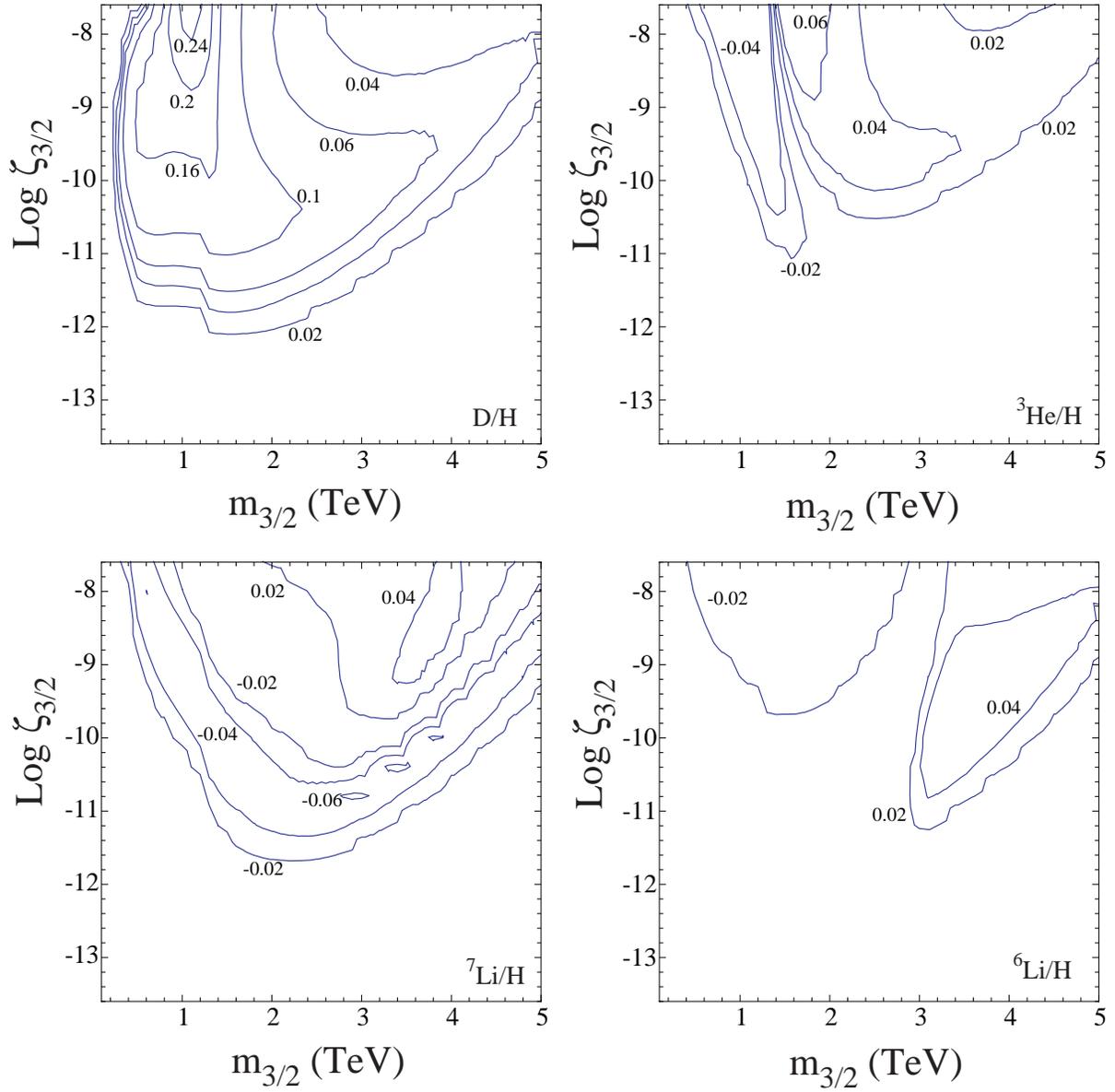,height=16cm}
\end{center}
\caption{\it
Similar to Fig.~\protect\ref{fig:reaction21}, for the reaction 23 ($n\he4 \rightarrow dnnp$).}
\label{fig:reaction23}
\end{figure}

We note that, although
the relative shift in the abundance is small for a 20\% variation in the rate, 
the \li7 abundance in this region (the diagonal strip where the 
abundance is decreased by 4-6\% in Fig.~\ref{fig:reaction21}) 
is already significantly reduced
when using the (unperturbed) non-thermal rates. 
In this region, several rates (principally reactions 21, 23, and 32) combine 
to lower the \li7 abundance when the gravitino abundance is sufficiently large.
However, subsequent variations in the rates do not make any further
significant changes in the abundances.  To help better understand this point
quantitatively, we show in Fig.~\ref{fig:livseps}, the \li7 abundance 
as a function of $\epsilon$ for rate 21 (other rates have $\epsilon_{i\ne21} = 0$) 
in the upper panel and as a function of $\epsilon_{21} = \epsilon_{23} = \epsilon_{32}$ in 
the lower panel.
As one can see, particularly in the latter case, when $\epsilon_{21,23,32} = -1$ and these
rates are shut off entirely, the abundance of \li7 is 4.2 $\times 10^{-10}$ (the 20\% 
decrease in \li7 is due to the remaining non-thermal reactions).  Furthermore, coherent 
variations in 
these rates of 20-40\% make relatively small changes in the abundances as reflected in 
Fig.~\ref{fig:reaction21}, and the effects of random variations in the rates would
clearly be smaller still.

Whilst reaction 23 ($n\he4 \rightarrow dnnp$) shows smaller variations in
\li7, the uncertainty (which we estimate at 40\%) is larger.  The effect on D/H is 
also pronounced, as seen in Fig. \ref{fig:reaction23}. The effect of reaction 22 
($n\he4 \rightarrow ddn$) is qualitatively similar but weaker.

We conclude this Section by summarizing the main results of our propagation of non-thermal
reaction rate errors into uncertainties in light-element abundances.
We find that the \he4 abundance is essentially unaffected by reaction rate errors.
For \li7, we find that no one reaction dominates the non-thermal
perturbations, which in turn means that errors in any given rate
only have a rather small (typically $\la 10\%$) effect on the \li7/H
abundance.  Non-thermal deuterium production is also not 
entirely controlled by
a single reaction, though the $n\he4 \rightarrow dnnp$ reaction 
clearly stands out as the most important, and the resulting errors
in D/H can go as high as $\ga 20\%$.  

These results have important implications
for our $\chi^2$ analysis, which, as we will see, is dominated by
\li7/H and D/H.  Since the \li7/H nuclear uncertainties are small
compared to the observational errors in the \li7/H abundance, the latter dominates
the lithium contribution to $\chi^2$.  Conversely, the D/H non-thermal
rate errors are significant in comparison to the observational errors,
and thus will have an important effect on the $\chi^2$ and ultimately
on the lithium problem.  From this we infer that the
reactions which most critically need improved nuclear data
are those which are important for deuterium production.

\section{$\chi^2$ Analyses of Benchmark CMSSM Scenarios}

To proceed with the $\chi^2$ analysis, we use Eq. (\ref{chi2}) to calculate $\chi^2$ for 
each point sampled in the  $(m_{3/2}, \zeta_{3/2})$ plane.  The reaction
rates are treated as
nuisance parameters and therefore, for each evaluation of 
$\chi^2$, each non-thermal rate is chosen from a Gaussian distribution about the 
mean rate with the uncertainty specified in the previous Section. At each point and
for each reaction considered, the 
difference between the rate chosen and its mean value, relative to the quoted uncertainty in the rate,
determines the corresponding $s_i$ in Eq. (\ref{chi2}).

From the results of the analysis in the previous Section, 
it is clear that it will be sufficient to focus on the effects of reactions 21, 23, and 32.
In principle, one could include all reactions in the $\chi^2$ analysis 
as nuisance parameters. However,
the inclusion of many more reactions would have only a marginal effect on lowering the 
$\chi^2$ contribution from the abundances, while at the same time increasing $\chi^2$ through $s_i^2$.  Since each rate typically increases $\chi^2$ by roughly one unit, one would need to 
gain at least one unit from the effect of the uncertainty in the rate on the element abundances.
Including the uncertainties of reactions beyond 21, 23, and 32 with finite sample sizes 
will typically lead to a 
larger value of $\chi^2$. 

\begin{figure}
\begin{center}
\epsfig{file=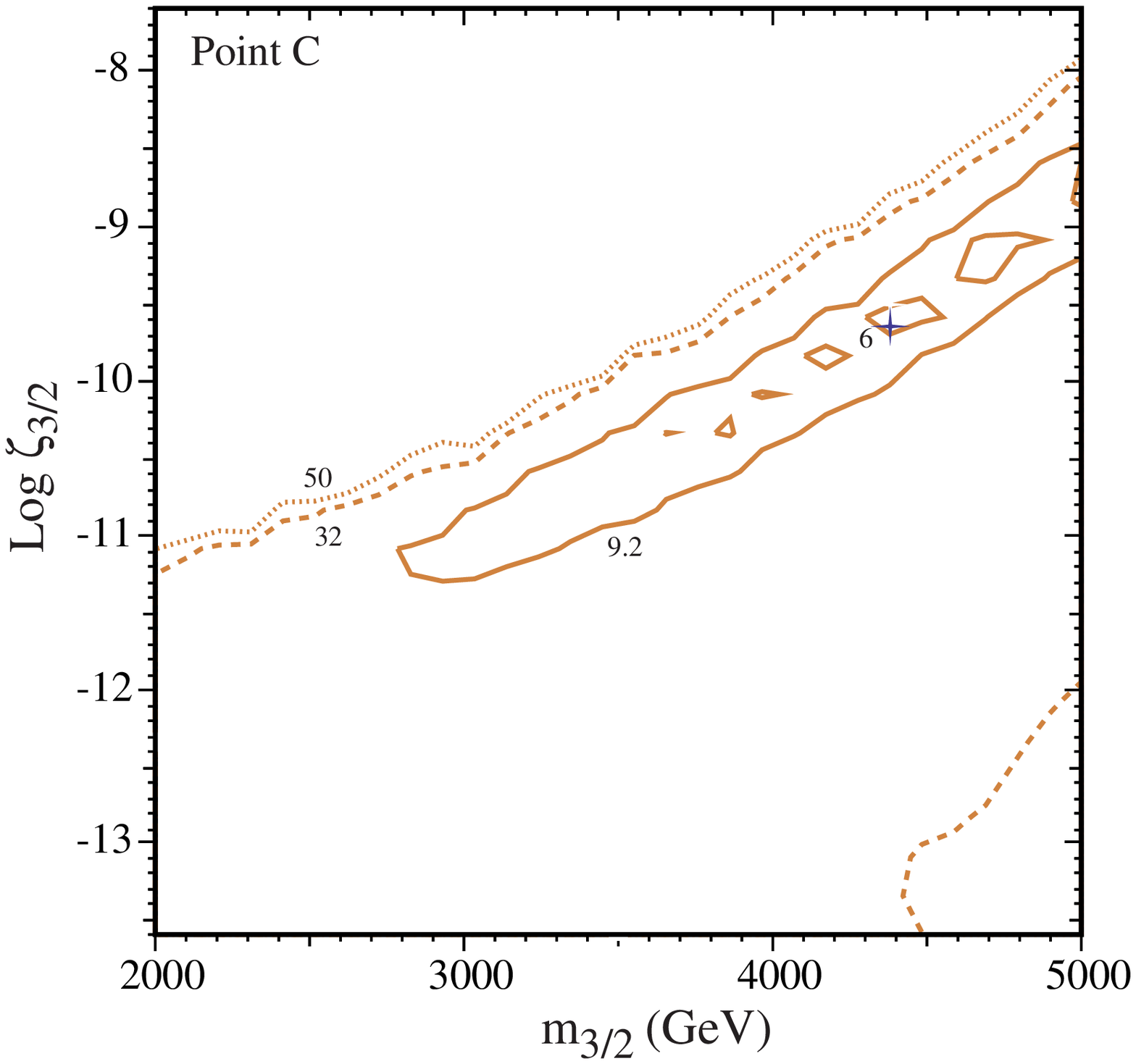,height=7cm}
\epsfig{file=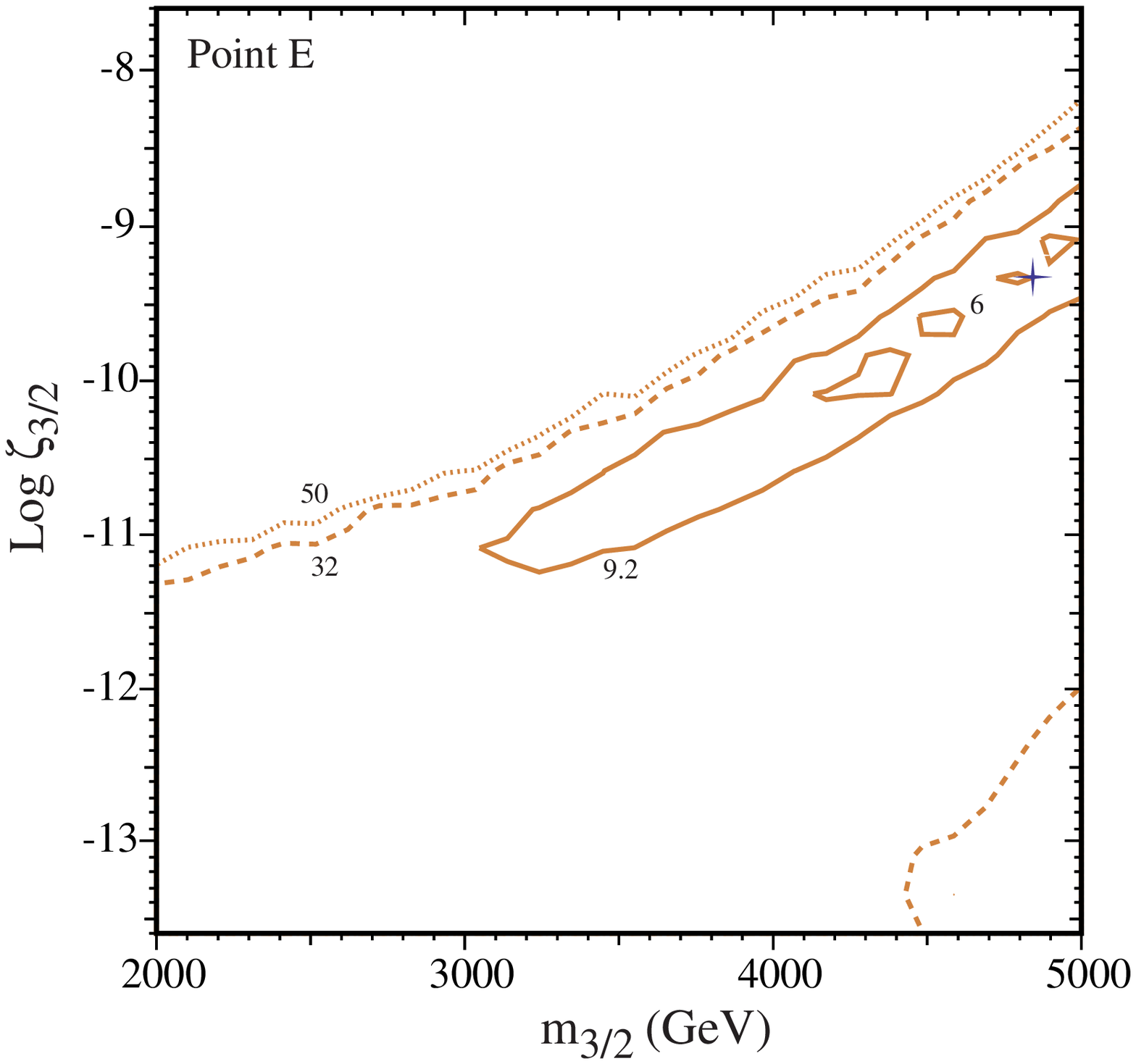,height=7cm}\\
\epsfig{file=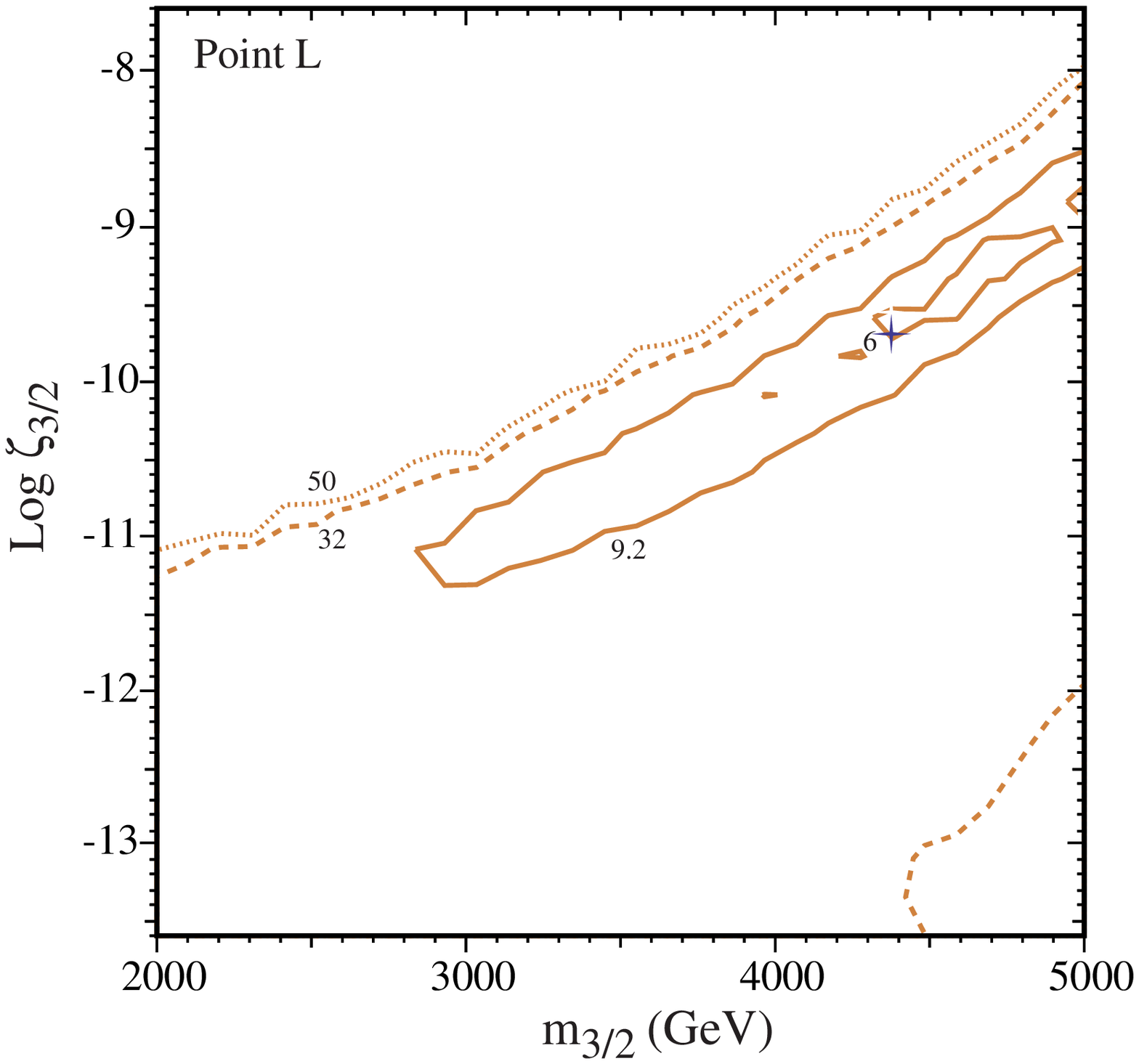,height=7cm}
\epsfig{file=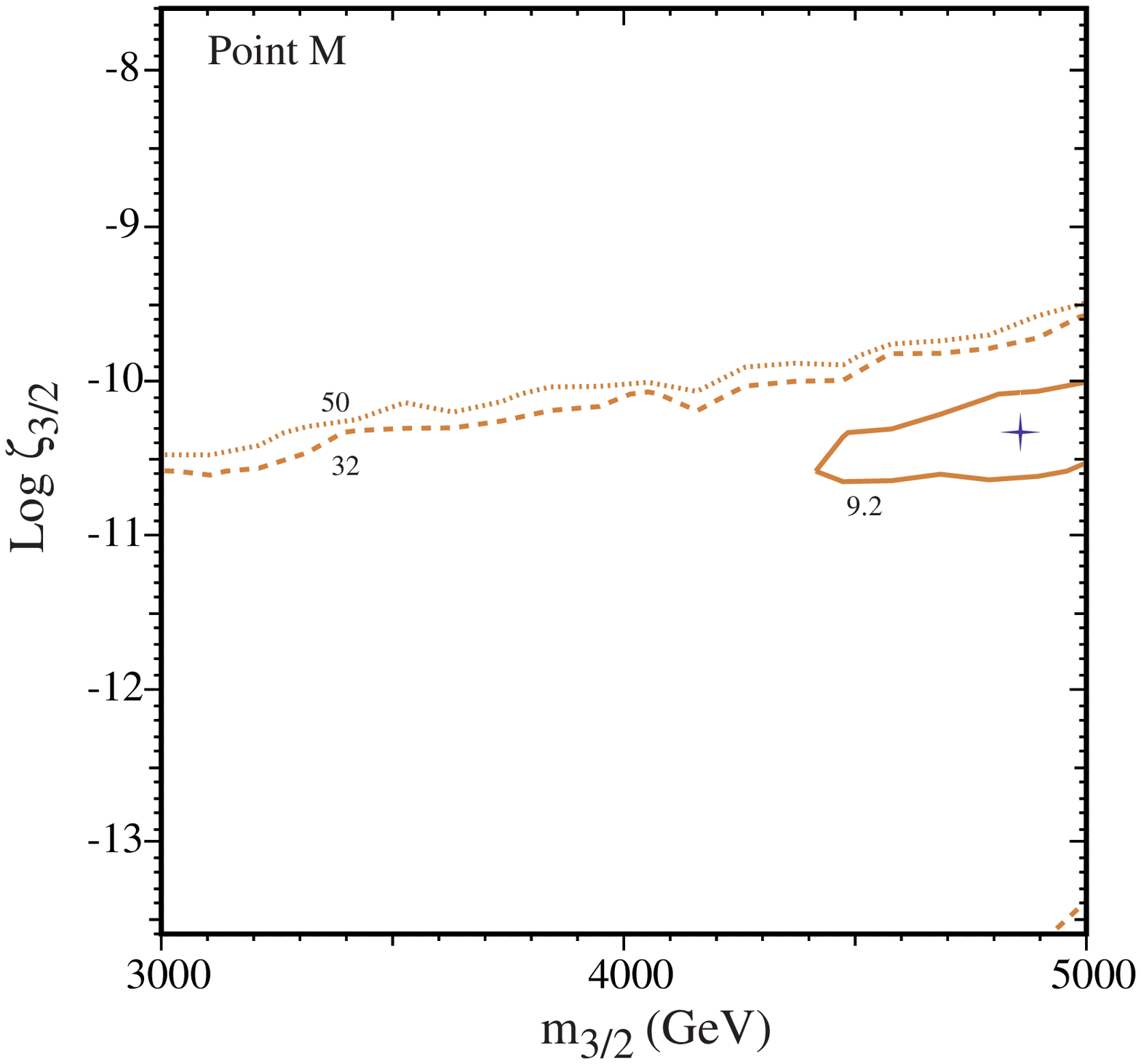,height=7cm}
\end{center}
\caption{\it
Contours of the $\chi^2$ function in the $(m_{3/2}, \zeta_{3/2})$ planes for the
benchmark CMSSM scenarios C (upper left), E (upper right), L (lower left) and
M (lower right), incorporating the uncertainties in the nuclear rates discussed in the text.}
\label{fig:scan}
\end{figure}

In~\cite{ceflos}, we discussed the application of the BBN constraints to
four benchmark CMSSM scenarios with specific values of the soft supersymmetry-breaking
parameters $m_{1/2}, m_0$ and $A_0$, $\tan \beta$, and the Higgs mixing parameter $\mu$,
labelled C, E, L and M~\cite{bench}. In each case, $A_0 = 0$ and $\mu > 0$ was chosen. 
The parameters corresponding to point C were given earlier.  For points E, L, and M
they are $(m_{1/2}, m_0, \tan \beta)$ = (300 GeV,1615 GeV, 10), (460 GeV, 310 GeV, 50), and 
(1840 GeV, 1400 GeV,  50) respectively.
Variants of these CMSSM scenarios with a massive gravitino
are characterized by the gravitino mass $m_{3/2}$ and its abundance $\zeta_{3/2}$.
The $(m_{3/2}, \zeta_{3/2})$ planes for benchmark scenarios C, E, L and M are shown in
Fig.~\ref{fig:scan}, displaying $\chi^2$ contours for the light-element abundances
calculated incorporating the nuclear reaction rate uncertainties discussed above.

In the limit of large $m_{3/2}$ and/or small $\zeta_{3/2}$, the value of the $\chi^2$
function approaches $\sim 31.7$, the same value as in standard BBN. This large
value of $\chi^2$ is due primarily to the \li7 problem. We see that in each of the
CMSSM scenarios in Fig.~\ref{fig:scan} there is a `trough' of much lower $\chi^2$
with a minimum at $\sim 5.5$, shown in each panel by a cross. 
We display contours of $\chi^2 = 6$ and 9.2, corresponding to the
95 and 99\% CLs for fitting to two parameters.  Also shown are the higher $\chi^2$ contours
of 32 (corresponding to the BBN value) and 50. 
We see that the $(m_{3/2}, \zeta_{3/2})$ planes are very similar for
benchmarks C, E and L. The plane for benchmark M is somewhat different, and
the minimum value of $\chi^2$ is slightly higher.
In Table \ref{tab:chi2}, we show the various abundances and $\chi^2$ contributions
for each of the three light elements for the standard BBN result and our best-fit point
for each of the four benchmark points. 

It is interesting to note the tension between
D and \li7.  At each of the best fit points, there is a considerable reduction in 
\li7, approaching the observational value.  
The minimum value $\chi^2 \sim 5.5$ certainly amounts to a `mitigation' of the \li7 problem, but
not a `solution', in the sense that since we are fitting two parameters ($m_{3/2}$ and $\zeta_{3/2}$)
and using 3 measurements, we have effectively only one degree of freedom and $\chi^2/{\rm d.o.f.} \sim 5.5$. 
However, this improvement  in \li7 comes at the
expense of D/H, which at this point begins to make a more significant contribution
to the total $\chi^2$.
On the other hand, the \he4
abundance $Y_p$ does not contribute significantly to the likelihood at any point in the parameter space. At the minimum, the deuterium
abundance contributes $\Delta \chi^2 \sim 1.5$, whereas the \li7 abundance
contributes $\Delta \chi^2 \sim 3.4$. Thus the previous 4- or 5-$\sigma$ \li7
problem is reduced to a $\la 2$-$\sigma$ problem. If this mitigation is to
lead to a complete solution, one or more of the nuclear reaction rates and/or measured
light-element abundances should lie outside its quoted uncertainty.

\begin{table}
\caption{Results for the best-fit points for CMSSM benchmarks C, E, L and M. 
The second set of results for C and M correspond to the globular cluster value for 
primordial \li7/H. The third and fourth entries for point C correspond to the higher
adopted uncertainty for D/H in field stars and to the globular cluster \li7 abundances, 
respectively.
\label{tab:best chi^2
table}}
\label{tab:chi2}
\begin{center}
\begin{tabular}{c c c c c c c c}
\hline\hline
 & $m_{3/2} \text{[GeV]}$ & $ \text{Log}_{10} (\zeta_{3/2} / \text{[GeV]})$ & $Y_p$ & D/H ($\times 10^{-5}$)& \li7/H ($\times 10^{-10}$) & $\sum s_i^2$ & $\chi^2$ \\
 \hline

 BBN & ------ & ------    & 0.2487 & 2.52 & 5.12 & ------ & 31.7 \\
\hline
 C   & 4380   & $-9.69 $  & 0.2487 & 3.15 & 2.53 & 0.26 & 5.5  \\
 E   & 4850   & $-9.27 $  & 0.2487 & 3.20 & 2.42 & 0.29 & 5.5  \\
 L   & 4380   & $-9.69 $  & 0.2487 & 3.21 & 2.37 & 0.26 & 5.4  \\
 M   & 4860   & $-10.29$  & 0.2487 & 3.23 & 2.51 & 1.06 & 7.0  \\
\hline
 C   & 4680   & $-9.39 $  & 0.2487 & 3.06 & 2.85 & 0.08 & 2.0  \\
 M   & 4850   & $-10.47$  & 0.2487 & 3.11 & 2.97 & 0.09 & 2.7  \\
\hline
 C   & 3900   & $-10.05$  & 0.2487 & 3.56 & 1.81 & 0.02 & 2.8  \\
\hline
 C   & 4660   & $-9.27$   & 0.2487 & 3.20 & 2.45 & 0.16 & 1.1  \\

\hline\hline
\end{tabular}
\end{center}
\end{table}

\begin{figure}
\begin{center}
\epsfig{file=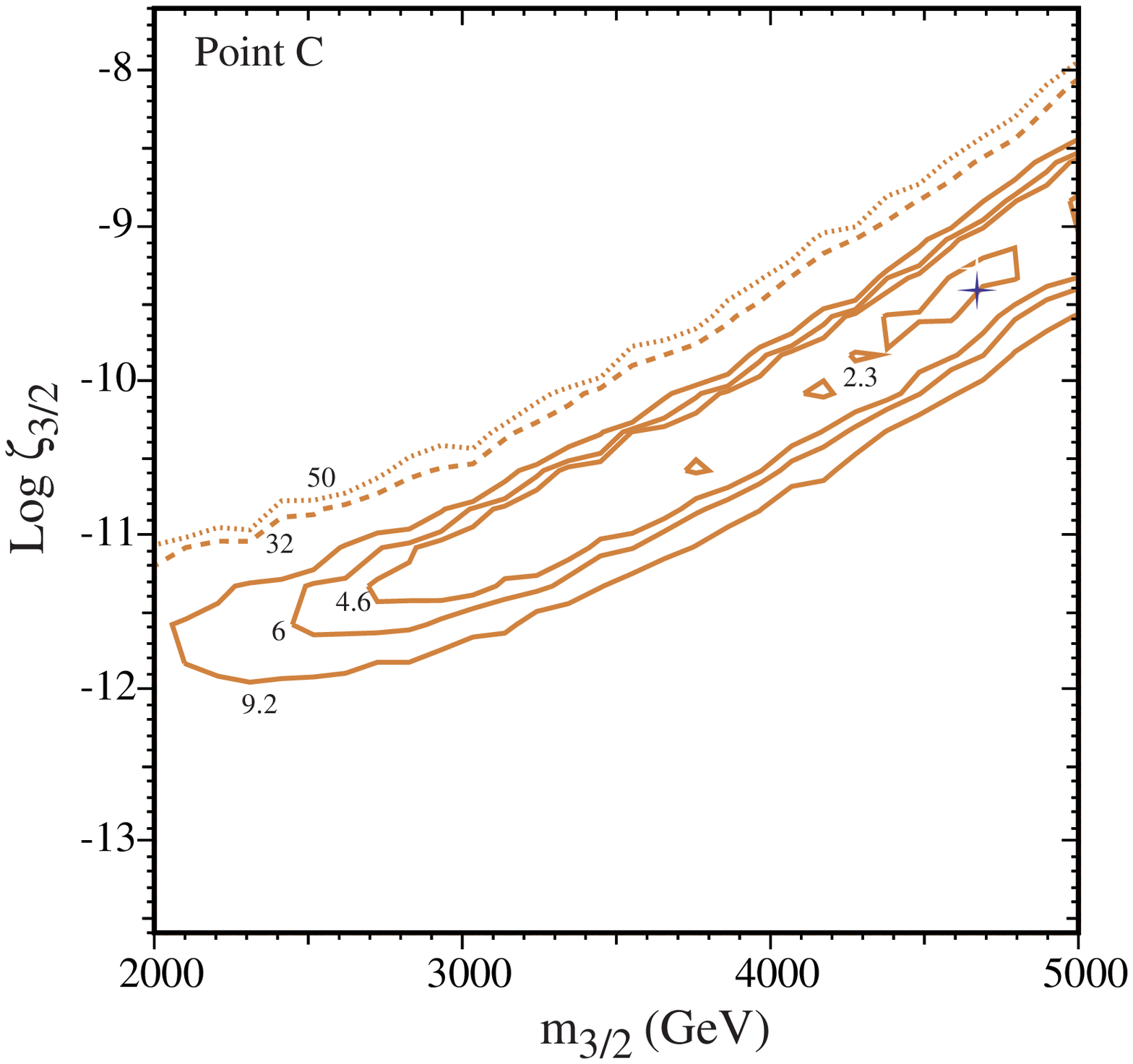,height=7cm}
\epsfig{file=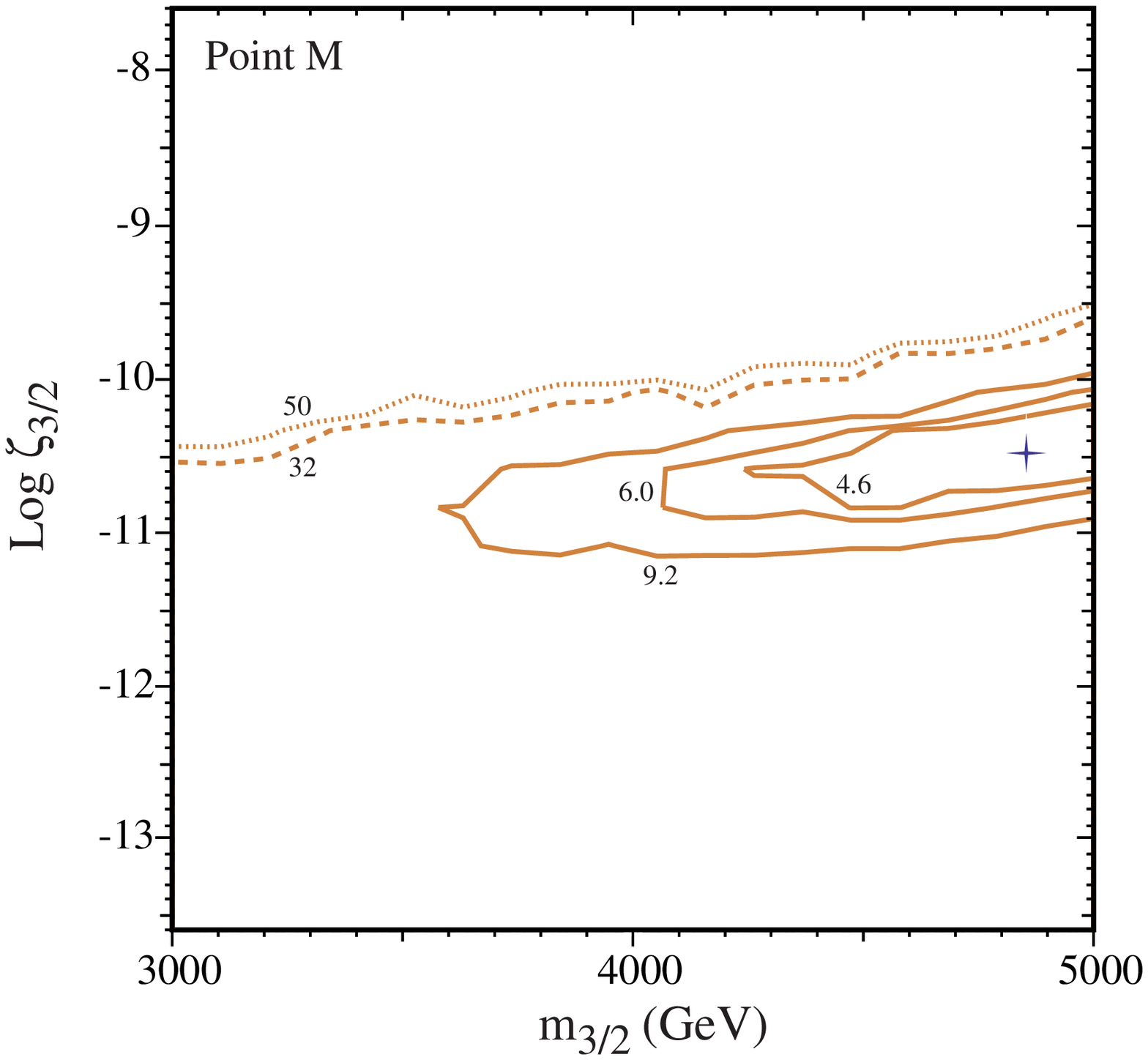,height=7cm}\\
\end{center}
\caption{\it
As in Fig.~\protect\ref{fig:scan}, contours of the $\chi^2$ function in the $(m_{3/2}, \zeta_{3/2})$ planes for the benchmark CMSSM scenario C (left) and M (right), assuming the globular cluster value of \li7/H.}
\label{fig:scancl}
\end{figure}

As an example,
in Fig. \ref{fig:scancl} we show the analogous results for the $\chi^2$ likelihood,
assuming the globular cluster value for \li7/H.  Results for this case for benchmark 
points C and M are also summarized in Table \ref{tab:chi2}. We now see the 
appearance of contours for $\chi^2$ = 4.6 and 2.3 corresponding to 
90 and 68 \% CLs respectively. The best-fit $\chi^2$ values 
drop considerably in this case, with values of 2.0 and 2.7 for points C and M respectively.
Thus a massive ($\ga 4$ TeV) gravitino can provide a potential solution of the 
lithium problem if globular cluster data is assumed to represent the primordial \li7 abundance.

\begin{figure}
\begin{center}
\epsfig{file=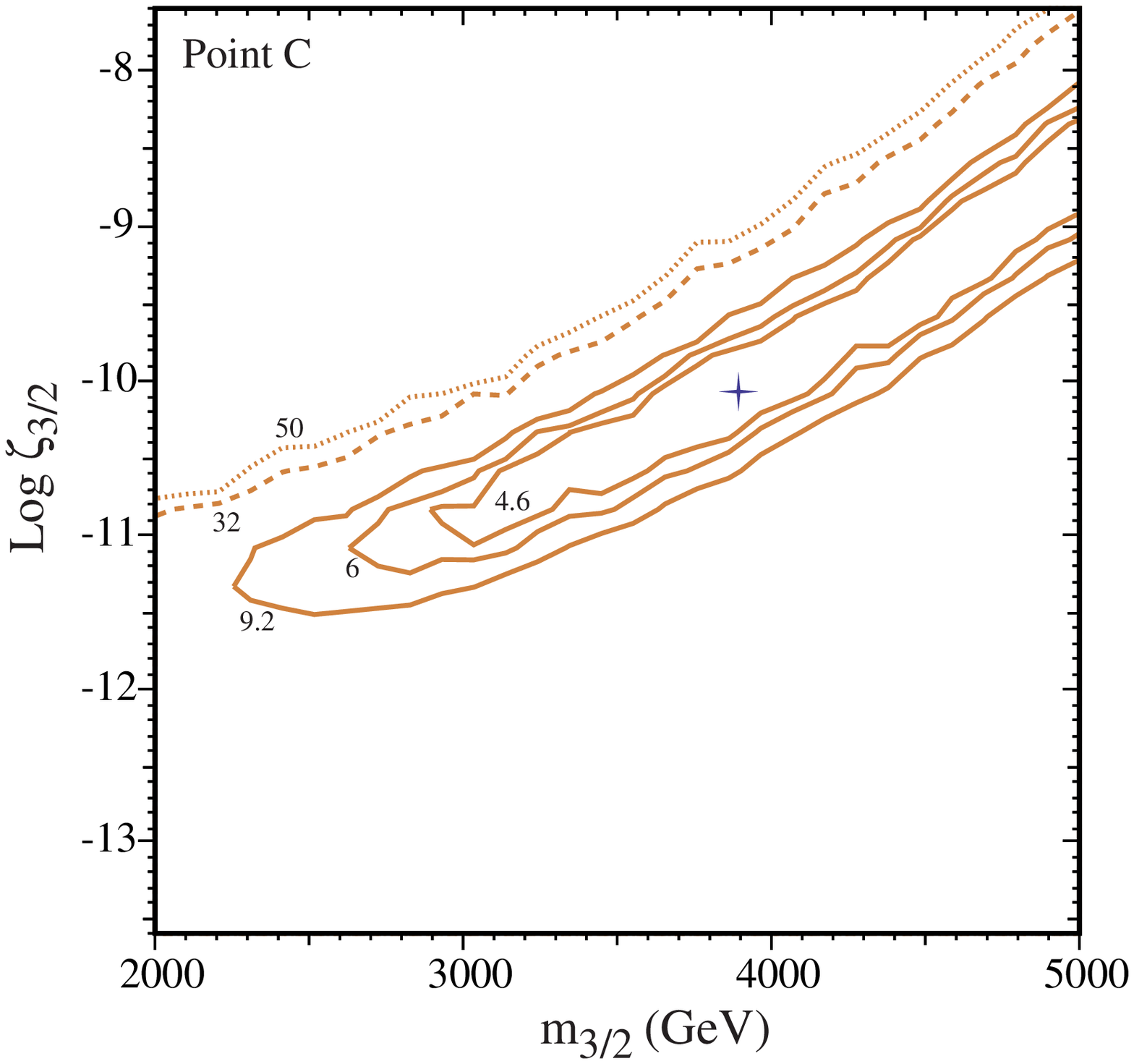,height=7cm}
\epsfig{file=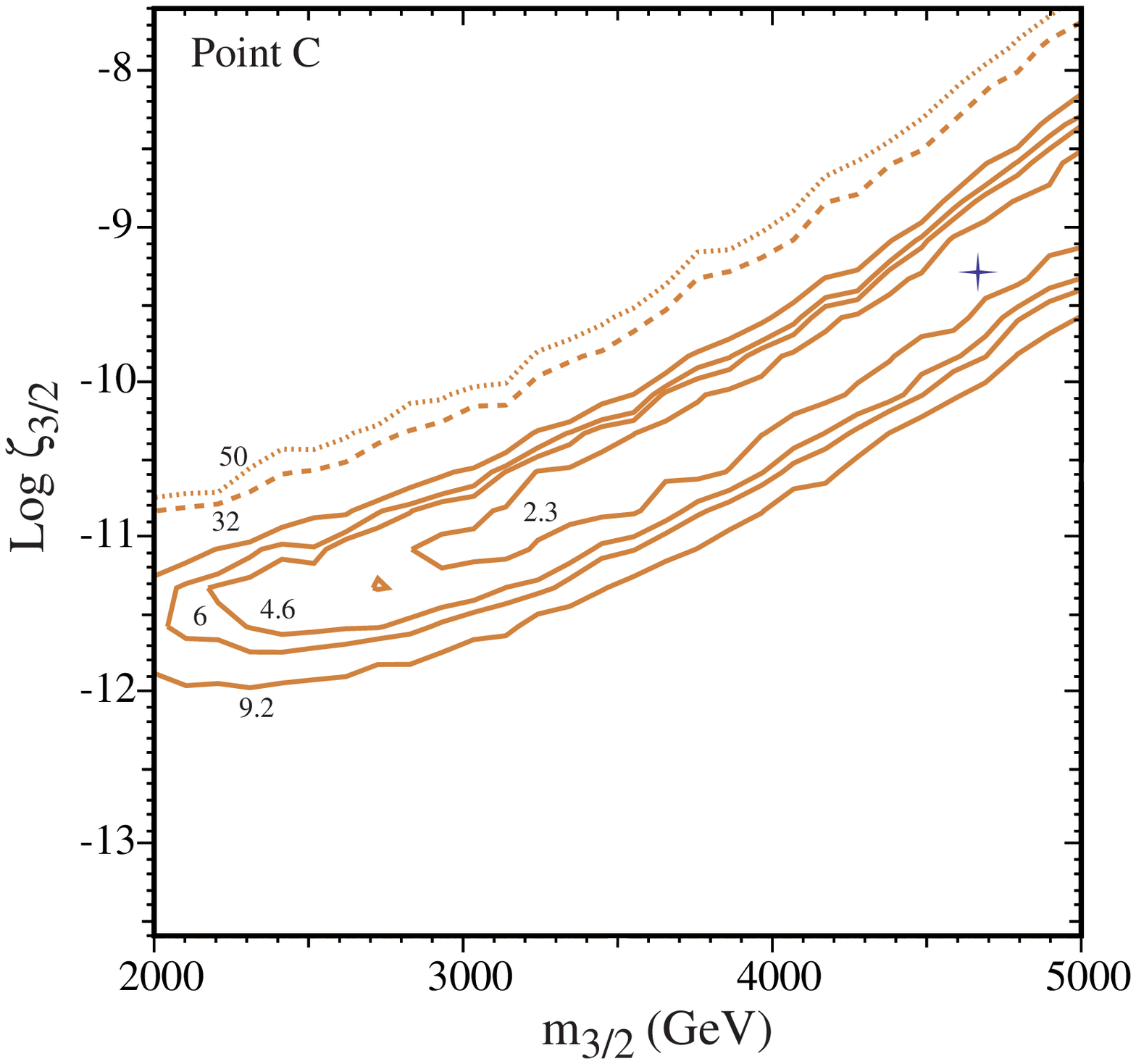,height=7cm}\\
\end{center}
\caption{\it
As in Fig.~\protect\ref{fig:scan}, contours of the $\chi^2$ function in the $(m_{3/2}, \zeta_{3/2})$ planes for the benchmark CMSSM scenario C assuming (left) a greater uncertainty in the 
observed D/H abundance and (right) also assuming the globular cluster value of \li7/H.}
\label{fig:scand}
\end{figure}

As discussed earlier, one may also consider the effect of increasing the size of the
uncertainty in the mean D/H abundance. Using an observed abundance of $(2.82 \pm 0.53)
\times 10^{-5}$, we obtain the $\chi^2$ contours seen in the left panel of
Fig.~\ref{fig:scand}, corresponding
to point C. In this case, we can obtain solutions with $\chi^2 = 2.8$ and a best-fit point with a \li7/H
abundance of  1.81 $\times 10^{-10}$ coming at the expense of a higher D/H abundance of 
$3.56 \times 10^{-5}$. When the globular cluster value of \li7/H is used together with the higher
D/H uncertainty, we can even find a best-fit solution with 
$\chi^2  = 1.1$: D/H = $3.20 \times 10^{-5}$ and \li7/H  = $2.45 \times 10^{-10}$, as seen in
the right panel of Fig.~\ref{fig:scand}.

\section{Summary and Conclusions}

We have presented in this paper an analysis of the modifications
of the cosmological light-element abundances that would be induced by the
late decays of massive particles, incorporating for the first time the uncertainties in
relevant nuclear reaction rates. We have analyzed the possible effects of the 36
different nuclear reactions shown in Table~\ref{tab:reactions}, 
and identified three as the most important, namely
$n\he4 \rightarrow npt$, $n\he4 \rightarrow dnnp$ and $n \he4 \rightarrow n n \he3$.

It is well known that there is a problem with the cosmological abundance of \li7 in
conventional BBN with no late-decaying particles, and a natural question is
whether this problem could be mitigated by some suitable late-decaying particle.
As an example of the possible applications of our uncertainty analysis, 
we have considered in this paper 
the late decays of massive gravitinos in various benchmark supersymmetric scenarios. 
It had been observed previously that there were regions of the parameter
spaces of these scenarios, corresponding to ranges of $m_{3/2}$ and $\zeta_{3/2}$,
where the cosmological \li7 problem might indeed be mitigated, and we have
made a likelihood analysis of this possibility incorporating uncertainties in the nuclear reaction
rates.

We confirm that there are indeed regions of the $(m_{3/2}, \zeta_{3/2})$
parameter planes in these scenarios where the global $\chi^2$ function is
reduced from its value $\sim 31.7$ in conventional BBN ($\sim 21.8$ if the
globular-cluster value for the \li7 abundance is adopted) to $\chi^2 \sim 5.5$.
This provides a very significant alleviation of the \li7 problem, reducing it
from a 4- or 5-$\sigma$ problem to a $\la 2$-$\sigma$ issue. The fact that
our best-fit points lie within the $\chi^2 = 6$ contours in Fig.~\ref{fig:scan}
implies that they have a goodness-of-fit slightly exceeding 5\%, which is
marginal for considering massive gravitino decay as a `solution' to the
cosmological \li7 problem.

The fact that this prospective solution exists for several 
choices of supersymmetric scenarios, with parameters that are 
relatively stable, suggests that it is a general feature of supersymmetric models.
For this potential solution to be confirmed, one or more of the following should
happen. 
\begin{enumerate}

\item
There might be some refinement in measurements of the
cosmological \li7 abundance leading to a shift in the central value and/or a change in the
assigned uncertainty. As we have shown in Fig. \ref{fig:scancl}, for
example, if the globular-cluster estimate of the \li7 abundance is adopted
(which would correspond to $\chi^2 \sim 21.8$ in standard BBN), the
decays of massive gravitinos could reduce $\chi^2$ to $\sim 2.0$.

\item
Alternatively, it is possible that the rates for one or more nuclear reactions
might lie outside the ranges favoured by the current measurements and their
assigned uncertainties. As we have pointed out, the measurements of some
of these non-thermal rates are sparse over the energy ranges of interest, and
improved coverage is certainly possible and desirable. The highest-priority
reactions for new cross-section measurements are $n\he4 \rightarrow npt$, 
$n\he4 \rightarrow dnnp$ and $n \he4 \rightarrow n n \he3$, which we have
shown to be the most relevant for this analysis. 

\item
Finally, we should mention
the possibility of some unidentified error in our analysis: we have given
reasons why we think its uncertainties are smaller than those mentioned
earlier in this paragraph, but we could be wrong.
\end{enumerate}

The supersymmetric possibility of `solving' the cosmological \li7 problem
is currently at a very intriguing stage. The decays of massive gravitinos are
one possibility, but it is a generic feature of supersymmetric theories with
gravity-mediated supersymmetry breaking that there is some late-decaying
massive particle, and other possible candidates exist. We plan to return
to some possibilities in a forthcoming paper. Clearly, this approach to
`solving' the cosmological \li7 problem would be given an enormous boost
if experimental evidence were to emerge for supersymmetry, either at the
LHC or in (in)direct searches for astrophysical dark matter. In this connection, we note that
there are good prospects for discovering supersymmetry at the LHC in many
benchmark scenarios, including points C, E and L discussed here~\cite{bench}, and
that there are also promising prospects for (in)direct detection of supersymmetric dark 
matter in scenarios C, E and L~\cite{EFFMO}. If supersymmetry were to be discovered,
the search for evidence of a possible metastable supersymmetric particle
would assume high priority, and it would be an exciting challenge to
correlate its possible roles in cosmology and in the laboratory. Then one
might indeed be justified in claiming that the cosmological \li7 problem was solved. 

\section*{Acknowledgments}
\noindent
The work of R.H.C.~was supported  by the U.S. National Science Foundation
Grants No.~PHY-01-10253 (NSCL) and Nos.~PHY-02-016783 
and PHY-08-22648 (JINA).
The work of K.A.O. and F.L. is supported in part by DOE grant DE-FG02-94ER-40823 at the 
University of Minnesota.
The  work of V.C.S. was supported by Marie Curie International Reintegration grant
ÒSUSYDM-PHENÓ, MIRG-CT-2007-203189.

\end{document}